# A compact butterfly-style silicon photonic-electronic neural chip for hardware-efficient deep learning

CHENGHAO FENG,[1,2,6], JIAQI GU[2,6], HANQING ZHU[2], ZHOUFENG YING[1,3], ZHENG ZHAO[1,4], DAVID Z. PAN [2,*], RAY T. CHEN[2,5,*]

[1]*Microelectronics Research Center, The University of Texas at Austin, Austin, Texas 78758, USA.*
[2]*Department of Electrical and Computer Engineering, The University of Texas at Austin, Austin, Texas 78705, USA.*
[3]*Alpine Optoelectronics, CA, USA.*
[4]*Synopsys Inc., CA, USA.*
[5]*Omega Optics, Inc., 8500 Shoal Creek Blvd., Bldg. 4, Suite 200, Austin, TX 78757, USA.*
[6]*These authors contributed equally.*
*chenrt@austin.utexas.edu, dpan@ece.utexas.edu

**Abstract:** The optical neural network (ONN) is a promising hardware platform for next-generation neurocomputing due to its high parallelism, low latency, and low energy consumption. Previous ONN architectures are mainly designed for general matrix multiplication (GEMM), leading to unnecessarily large area cost and high control complexity. Here, we move beyond classical GEMM-based ONNs and propose an optical subspace neural network (OSNN) architecture, which trades the universality of weight representation for lower optical component usage, area cost, and energy consumption. We devise a butterfly-style photonic-electronic neural chip to implement our OSNN with up to 7× fewer trainable optical components compared to GEMM-based ONNs. Additionally, a hardware-aware training framework is provided to minimize the required device programming precision, lessen the chip area, and boost the noise robustness. We experimentally demonstrate the utility of our neural chip in practical image recognition tasks, showing that a measured accuracy of 94.16% can be achieved in hand-written digit recognition tasks with 3-bit weight programming precision.

## 1. Introduction

Deep neural networks (DNNs) have demonstrated superior performance in various intelligence tasks, such as image recognition, decision making, and language translation [1–4]. Hardware accelerators capable of performing high-speed, energy-efficient, and parallel multiply-accumulate (MAC) operations are in high demand with the rapidly-escalating DNN model size and data volume. However, electronic digital hardware accelerators, including but not limited to graphical processing units (GPUs), field-programmable gate arrays (FPGAs) [5], and other digital application-specific integrated circuits (ASICs) [6], are inevitably limited by millisecond-level latency, high energy consumption, excessive heat, and high interconnect cost [7,8]. In contrast, analog neuromorphic computing represents a paradigm shift in efficient DNN acceleration, significantly increasing parallelism and energy efficiency [9,10].

The optical neural network (ONN) is a promising analog artificial intelligence (AI) accelerator that features low latency, wide bandwidth, and high parallelism of light [11–15]. Earlier work has presented a range of high-performance integrated photonic neural networks that implement multi-layer perceptrons (MLPs) [11,16,17] or convolutional neural networks (CNNs) [18,19]. The fundamental matrix-vector multiplication (MVM) unit is realized using Mach-Zehnder interferometer (MZI) arrays or microring-resonator (MRR) arrays. By tuning the phase shifters in MZIs or the transmission of MRRs, these photonic systems are designed to implement universal linear operations or general matrix multiplication (GEMM) with a relatively high requirement in device control precision. Recent studies show that the construction of DNNs can move beyond conventional GEMM with restricted matrix parameter space, e.g., low-rank NNs [20–22] and structured NNs [23–25], which shows not only considerable hardware efficiency improvement but also comparable representability to classical GEMM-based NNs. We refer to such NN architectures as subspace neural networks. The success of such a design concept can be reproduced in ONNs by trading the universality of weight representation for higher hardware efficiency. Several structured ONNs have been proposed to reduce the number of optical components, e.g., the fast-Fourier-transform-based (FFT-based) ONN [26–28]. In this work, we further explore this subspace NN design concept in the optical domain and experimentally demonstrate a novel butterfly-style photonic-electronic neural chip (BPNC) with superior hardware efficiency and compactness.

Additionally, noise-tolerant ONN training currently still lacks an efficient, scalable, and physically-evaluated solution. As is the case with other analog computing platforms, ONNs will inevitably encounter performance degradation or even malfunction due to non-ideal factors, e.g., process variations [29,30], limited control precision [31,32], and dynamic noises [33]. Recently, on-chip training has become an appealing trend towards noise-resilient ONNs. Numerous on-chip training algorithms have been proposed to directly optimize optical devices with in-situ noise handling [34–39]. However, prior ONN on-chip training protocols suffer from algorithmic inefficiency and require costly hardware overhead, e.g., phase detection [34], high-resolution optical component control [35], or per-device field monitoring [34,39]. Therefore, applying them to practical ONN training is still technically challenging.

In this work, we propose an OSNN for next-generation hardware-efficient deep learning. Our proposed OSNN partitions each layer's weight matrix into smaller $k \times k$ ($k$=4,8) submatrices with restricted parameter space. Our architecture can achieve photonic neural computing with $7\times$ fewer trainable optical components compared to MZI-based ONN architectures designed for general MVMs [11], resulting in a $3.3\times$ smaller footprint and $5.5\times$ lower latency. The number of trainable optical components can be further reduced by ~70% using structured circuit pruning [40] with negligible (<0.2%) task performance loss. Moreover, an efficient and scalable hardware-aware training framework is experimentally deployed to enable ONN training with high noise robustness and low control precision requirement. Our OSNN is then experimentally demonstrated on a 4×4 BPNC and evaluated on the MNIST hand-written digits classification task [41] with a measured accuracy of 94.16%. Our performance analysis reveals that our OSNN can achieve a computational density of ~225 tera (10E+12) operations per second/mm$^2$ (TOPS/mm$^2$) and energy efficiency of ~9.5 TOPS/W using compact optical devices, e.g., microdisk-based active devices [42]. Our proposed OSNN architecture and hardware-aware training framework provide a synergistic solution that unleashes the power of optics from a novel co-design perspective and pushes the limits of next-generation efficient AI.

## 2. Optical subspace neural network

The proposed OSNN and its training framework are depicted in Fig. 1(a). The mathematical representation of one layer in a typical DNN with $n$ inputs and $m$ outputs is shown in Fig. 1(b), along with its hardware implementation shown in Fig. 1(c). Here we partition the $m \times n$ weight matrix $\boldsymbol{W}$ into $m^* \times n^*$ submatrices $\{\boldsymbol{W}_{i,j} \in \mathbb{C}^{k \times k}\}_{i \in [n^*], j \in [m^*]}$. The input vector $\boldsymbol{x_{in}}$ is encoded as the amplitude of the optical signals and will also be partitioned into $n^*$ segments $\boldsymbol{x_{in}} = (\boldsymbol{x_{in}^1}, \boldsymbol{x_{in}^2}, \dots, \boldsymbol{x_{in}^{n^*}})$. Thus, the MVM operation can be expressed using the block matrix multiplication formula as follows,

$$x'_{out} = W x_{in} = \begin{pmatrix} \sum_{j=1}^{n^*} W_{1,j}\, x_{in}^j \\ \sum_{j=1}^{n^*} W_{2,j}\, x_{in}^j \\ \vdots \\ \sum_{j=1}^{n^*} W_{m^*,j}\, x_{in}^j \end{pmatrix}. \qquad \text{Eq. (1)}$$

Each submatrix $W_{i,j}$ can be decomposed as $W_{i,j} = B\Sigma_{i,j}P$, where $B$ and $P$ are both $k \times k$ unitary matrices shown in Fig. 1(d), and $\Sigma_{i,j}$ is a $k \times k$ diagonal matrix. Here we use two butterfly-style programmable photonic integrated circuits (PICs), namely, a butterfly-style transform unit ($B$ unit) and a projection unit ($P$ unit), to implement the unitary matrices $B$ and $P$ using phase shifters, directional couplers, and waveguide crossings, while the diagonal matrix unit ($\Sigma$ unit) is composed with a column of modulators. Fig. 1(d) and Fig. 1(e) show the photonic circuit structure of these matrix units when $k = 4$ and $k = 8$, respectively.

One of the key advantages of our OSNN is that the chip footprint and the total number of trainable optical devices are considerably smaller than previous GEMM-based MZI-ONN. Specifically, while an $k \times k$ MZI array consumes $\mathcal{O}(k^2)$ MZIs, our $B$ and $P$ units only use $\mathcal{O}(k \log_2 k)$ couplers and phase shifters. Besides, instead of having all devices to be trainable, only the $\Sigma_{i,j}$ units need to be trained. The $B$ and $P$ units will not be modified throughout the training and mapping processes after their desired states are accomplished by tuning the phase shifters in them. As a result, the total number of trainable optical devices is $\frac{mn}{k}$ in an $n$-input, $m$-output layer, significantly reducing the weight loading cost and reprogramming complexity.

Based on the statistical evaluation [43], our butterfly-style $B\Sigma P$ block demonstrates good flexibility and matrix expressivity by only using $1/k$ total trainable components compared to MZI arrays (details in Supplementary Note 1). The $B$ and $P$ units in OSNN can flexibly support a wide range of unitary transforms. For instance, when $k = 4$, our butterfly unit $B$ itself can express 80.2% arbitrary unitary matrices, and the $B\Sigma P$ block can realize 64.4% fidelity in expressing general matrices. As depicted in Fig. 1(d), several commonly used structured matrices can be realized by configuring the phase shifters in $B$ and $P$ units. For example, a block-circulant matrix can be realized (Fig. 1(d(1)) when the $P$ unit performs optical FFT while the $B$ unit performs optical inverse FFT (IFFT) [23]. Furthermore, our $B$ and $P$ units can realize Hadamard transformation (HT), which is a popular choice to construct efficient DNNs [44]. Detailed proves can be found in supplementary Notes 1 and 2. Figure 1(d(2)) shows the matrix pattern when both $P$ and $B$ units implement HT. The superior versatility and expressivity of our $B\Sigma P$ units guarantee that our OSNN can have enough learning capability.

Another essential property of our OSNN is that different $\Sigma_{i,j}$ units can share the $B$ and $P$ units, leading to significant chip area reduction. Since all $W_{i,j}$s are constructed by the same $B, P$ transforms, $B$ and $P$ units can be reused in the optical domain. Here, we rewrite Eq. 1 with matrix multiplication's distributive and associative properties:

$$x'_{out} = W x_{in} = \begin{pmatrix} \sum_{j=1}^{n^*} B\Sigma_{1,j} P x_j \\ \sum_{j=1}^{n^*} B\Sigma_{2,j} P x_j \\ \vdots \\ \sum_{j=1}^{n^*} B\Sigma_{m^*,j} P x_j \end{pmatrix} = \begin{pmatrix} B\sum_{j=1}^{n^*} \Sigma_{1,j} \aleph_j \\ B\sum_{j=1}^{n^*} \Sigma_{2,j} \aleph_j \\ \vdots \\ B\sum_{j=1}^{n^*} \Sigma_{m^*,j} \aleph_j \end{pmatrix}, \qquad \text{(Eq. 2)}$$

where $\aleph_j = P x_j$. By sharing the unitary matrix units, one can implement an MVM operation of size $m \times n$ with only $m/k$ $P$ units and $n/k$ $B$ units, dramatically reducing the footprint of the OSNN compared to previous FFT-based ONN architecture, which requires $\frac{mn}{k}$ $P$ and $B$ units [26]. The mechanism of the optical architecture shown in Fig. 1(b) can then be stated as follows: First, the input optical signal $x_{in}$ is partitioned into $n^*$ segments $x_j$ s and then propagate through the $\boldsymbol{P}$ units to generate $\aleph_j$s, which will then be distributed via a fanout network to $m^* \times n^*$ diagonal matrix units $\boldsymbol{\Sigma}_{i,j}$s. After propagating through the $\boldsymbol{\Sigma}_{i,j}$ units, the signals will be combined and fed into each $\boldsymbol{B}$ unit with a combiner network to obtain the MVM result. Finally, as with all typical NN architectures, a nonlinear activation unit ($\boldsymbol{\sigma}$ unit) is required to generate the output of the layer $\boldsymbol{x_{out}}$, which has been realized by all-optical non-linear devices or optoelectronic circuits in previous work [45,46]. In this work, we assume the activation functions are realized electronically. We omit details on this and do not consider its effects on system performance in our discussion hereinafter.

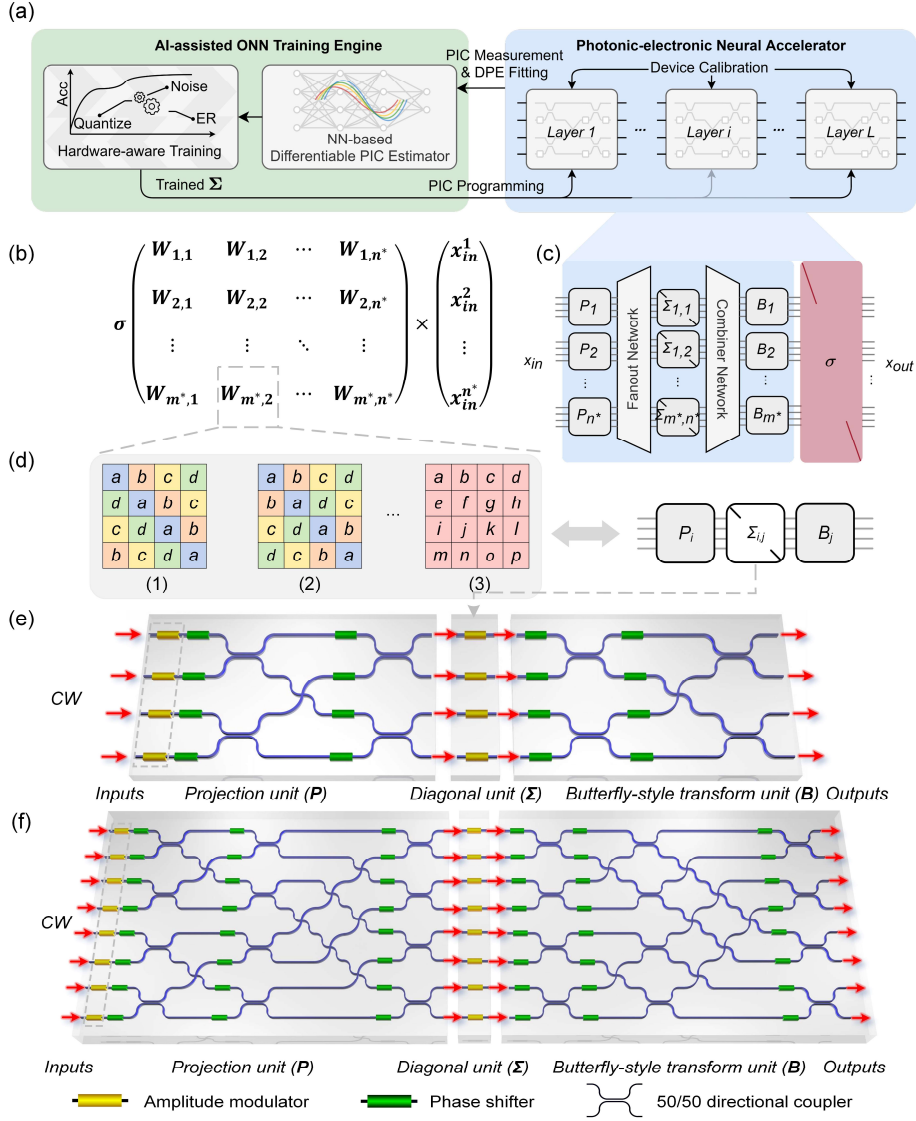

**Fig. 1 General architecture of the optical subspace neural network (OSNN).** The hardware-aware training framework is shown in (a). The mathematical representation of one layer in our OSNN is shown in (b), where an $m \times n$ matrix is partitioned to $\frac{mn}{k^2}$ blocks. The hardware representation of one layer is shown in (c), which consists of $n^* = \frac{n}{k}$ projection units ($P$ units), $m^* = \frac{m}{k}$ butterfly-style transform units ($B$ units), $m^* \times n^*$ diagonal matrix units ($\Sigma$ units). The fanout network and the combiner network are used to distribute and combine the optical signals to different optical paths. By setting phase shifters in $B$ units and $P$ units, several popular structured ONNs based on (1) Fast Fourier transform and inverse Fast Fourier transform, (2) Hadamard transform, or (3) other implementable transforms are shown in (d). (e) Schematic of 4×4 $B$, $P$, and $\Sigma$ units, respectively, which are the building blocks of a 4-point OSNN. (f) Schematic of 8×8 $B$, $P$, and $\Sigma$ units, respectively, which are the building blocks of an 8-point OSNN. To train the OSNN, we first perform an on-chip device calibration on our OSNN. The measurement data are learned and modeled in our hardware-aware training flow. Besides, other factors such as the precision of controlling signals, limited extinction ratio (ER) of input modulators, and other noises are also considered in our training engine to improve the accuracy and robustness of our OSNN.

## 3. Hardware-aware training framework

After manufacturing, fabrication variances in optical components and dynamic noises will introduce uncertainty into the ONN. Moreover, the numerical resolution of implementable weight matrices is limited by the precision of electrical signals used to program the ONN. Consequently, the task performance will deteriorate. To remedy this robustness issue, we build a multi-stage hardware-aware training framework.

The general procedures of the training framework are summarized in Fig. 1(a). First, the OSNN is calibrated on-chip to measure the performance of tunable components and prepare the state of the photonic chip to approach the desired transfer matrix. In reality, the actual transfer matrix of the photonic neural chip deviates from the designed one due to performance variations of the optical components, e.g., the unbalanced splitting ratio of directional couplers. Second, to model the non-ideal behavior and predict the response of the real optical neural chip, we develop an NN-based differentiable PIC estimator (DPE) using measurement data and AI algorithms. Our DPE explicitly models the behavior of the real physical chip during forward and gradient backpropagation that enables gradient-based physical-variation-aware optimization. The third step involves determining the DNN parameters and mapping them to the electrical control signals using a hardware-aware training and parameter mapping process. The DPE is used to efficiently emulate the real chip response to enable variation-aware gradient backpropagation. Quantization-aware training with dynamic noise injection techniques is used to improve the noise tolerance with limited device control resolution. Thus, our OSNN can achieve the performance target despite control precision restrictions and other non-idealities. More details of our hardware-aware training framework are shown in Supplementary Note 3. Compared to prior on-chip training protocols based on derivative-free optimization algorithms [36,47–49] and gradient approximation using ideal simulation models [50], our AI-assisted ONN learning shows considerably higher scalability and effectiveness in robust optical neural chip training.

## 4. Experiment

In this work, we experimentally demonstrate the practicality of the OSNN on the silicon photonics platform using a butterfly-style photonic-electronic neural chip (BPNC) capable of implementing 4×4 $B\Sigma P$ blocks in our OSNN. The layout of the chip was drawn and verified using Synopsys OptoDesigner, while the chip was fabricated by the Advanced Micro Foundry (AMF). The schematic of the BPNC is shown in Fig. 2(a), while the close-ups of its components, such as phase shifters, 50-50 directional couplers, and crossings, are depicted in Fig. 2(b). The unitary matrix units $B$/$P$ are marked in red/green in Fig. 2(a). The active phase shifters in these regions support enough flexibility to realize different unitary transforms but note that they are not optimized as parameters during ONN training. The diagonal matrix unit ($\Sigma$ unit) is built using an array of MZI attenuators for magnitude and phase control [51].

The schematic of the testing setup is shown in Fig. 3. Continuous-wave (CW) light of different wavelengths is coupled in different input grating couplers separately. There are three reasons to use multi-wavelength inputs in our BPNC: First, we can use compact resonator-based modulators as input modulators and avoid additional hardware costs for phase control, which requires high-speed phase shifters. Second, phase detection at the outputs can be avoided using multi-wavelength inputs[51]. Third, multi-wavelength inputs eliminate the phase fluctuations of optical signals in off-chip fibers and improve the robustness of OSNN to input phase noises, which have been reported in other work[53]. More details about the weight matrix of the BPNC when we use multi-wavelength inputs are provided in Supplementary Note 4. Using multi-wavelength inputs, our BPNC can express arbitrary non-negative 4×4 matrices with a surprisingly high fidelity of 92.2% (See Supplementary Note 1). The input modulators and phase shifters of the BPNC are programmed by a high-precision multi-channel digital-to-analog converter (DAC). Off-chip photodetector arrays will collect the output signals, which will subsequently be read using oscilloscopes or analog-to-digital converters (ADCs). A microcontroller is used to write electrical signals to the DAC and read the output signals in this work. The measurement data are processed by computers to train and implement the DNN model. It should be noted that current fabrication and packaging technologies enable the integration of electrical circuits, photodetectors, and the laser on a single chip [52] with potentially much higher compactness, shorter interconnect paths, and higher efficiency. The experimental setup is described in detail in Supplementary Note 5.

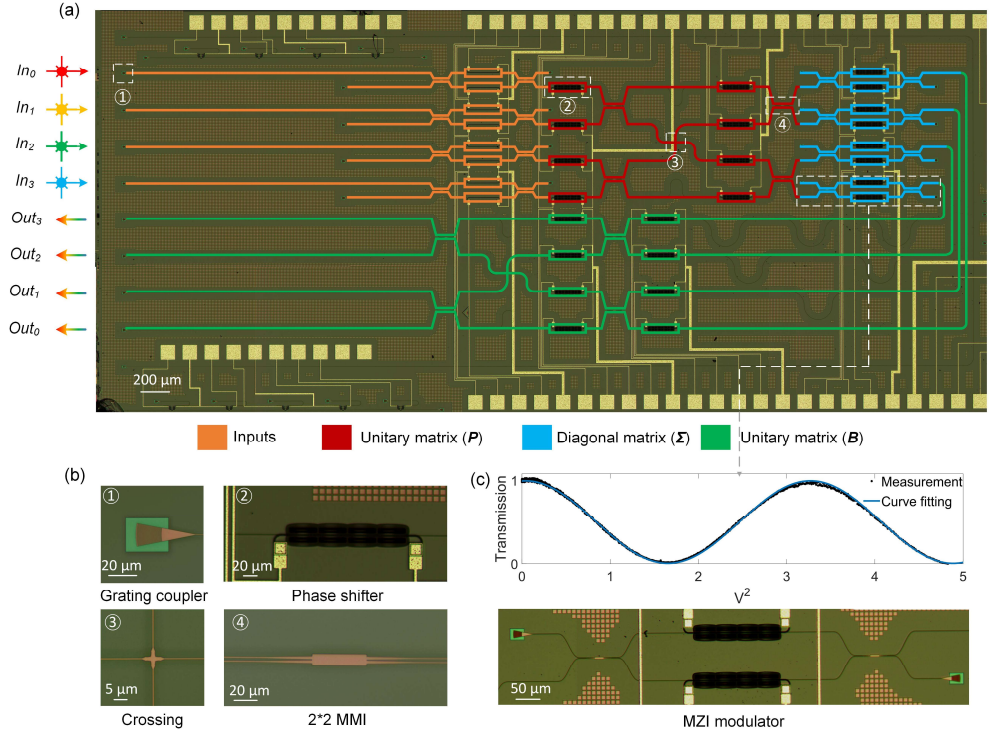

**Fig. 2 Schematic of the butterfly-style silicon photonic-electronic neural chip.** The micrograph of the neural chip is shown in (a). The input optical beams with different wavelengths are shown in different colors. The necessary optical components are highlighted in (b). (c) shows the schematic and the normalized transmission curve of an MZI attenuator in the diagonal matrix unit ($\Sigma$ unit). Only the attenuators in $\Sigma$ are programmed in training.

Here we experimentally implement the multi-stage hardware-aware training flow on our BPNC. In the calibration stage, the performance of modulators and phase shifters in the BPNC are first calibrated individually, such that we can precisely control the state of active devices, especially the input modulators and the $\Sigma$ matrix (The calibration results are detailed in Supplementary Note 6). The second stage is to learn desired device configurations via ONN training. We first program the BPNC with representative input signals and phase shifter control voltages and collect the corresponding outputs. The above measured input-output pairs are used to train our differentiable PIC estimator for accurate and efficient chip response modeling. Then, we embed our DPE into our ONN training procedure to effectively enable hardware-aware training. Quantization-aware training and dynamic noise injection techniques are used during training to adapt the ONN model to limited phase shifter control resolutions and boost the PIC robustness to dynamic system noises.

In this work, we construct a CNN with our BPNC and benchmark its performance on a hand-written digits classification dataset MNIST [41]. We use MVM operations to implement CNNs with a widely-applied tensor unrolling method (im2col) [53], as detailed in

Supplementary Note 7. Figure 4(a) illustrates the network structure. Here, large-size tensor operations are partitioned into 4×4 blocks and mapped onto our BPNC. When the voltage control resolution is set to 3-bit (8 attenuation levels for each MZI attenuator in the $\Sigma$ unit), the inference accuracy of the CNN reaches 94.16% in our experimental demonstration, comparable to the simulated value of 94.59%. The confusion matrix depicting the prediction results is shown in Fig. 4(b). Figure 4(c) visualizes tested output images after being convolved by learned kernels. Figure 4(d) shows the tested probability distribution of different hand-written digits. More testing results are included in Supplementary Note 8, where we evaluate the accuracy of OSNN with different control voltage ranges and control resolutions. They will also be discussed in the following section.

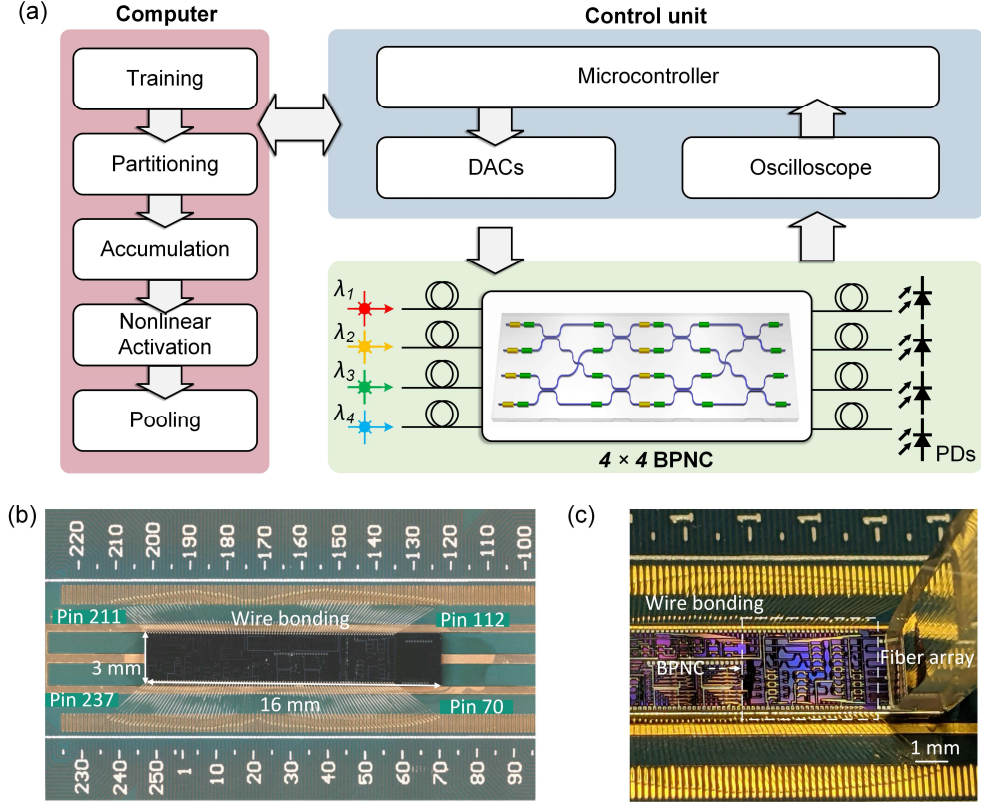

**Fig. 3 Experimental setup of OSNN.** (a) Schematic of our OSNN test flow. The entire MVM is first partitioned into multiple 4×4 blocks, and each block is implemented optically on a butterfly-style photonic-electronic neural chip (BPNC). (b) shows the wire-bonded photonic chip and its starting/ending electrical pin numbers, while (c) is the photography of the chip testing setup. The parameters and the input signals are programmed by a multi-channel digital-to-analog converter (DAC), while the output signals are read by the oscilloscope. Both the oscilloscope and the DAC are controlled by a microcontroller. The MVM results are provided to the computer for data processing in order to train and deploy the DNN.

## 5. Discussion

**Footprint.** Our OSNN outperforms SVD-based MZI ONN architectures [47] in the number of trainable devices and the footprint. Rather than deploying area-costly MZI arrays, we use basic optical components such as directional couplers, phase shifters, and crossings to construct the unitary matrix units $\boldsymbol{B}$ and $\boldsymbol{P}$. The second reason that leads to our superior compactness is that many $\Sigma$ units share the $\boldsymbol{B}$ and $\boldsymbol{P}$ units, which reduces the chip area for implementing unitary transforms. Shown in Fig. 5(a), when the matrix size is 32×32, our 8-point OSNN consumes ~3.6× fewer phase shifters and ~4.8× fewer directional couplers, leading to ~3.3× footprint reduction compared to SVD-based MZI ONN architectures [47] with the same matrix size and optical component selection. Footprints of different ONN architectures are estimated by summing the areas of their constituent optical components provided by the same foundry (AMF). See the detailed evaluation of the chip area in Supplementary Note 9.

The chip area or hardware cost of OSNN can be further optimized with structured circuit pruning strategies. In an $n$-input, $m$-output layer, the $\frac{mn}{k^2}$ diagonal matrix units can be treated as $\frac{mn}{k^2}$ parameter groups. When all of the transmission coefficients in one $\pmb{\Sigma}$ unit are zeros, this unit or parameter group is unnecessary and can be omitted in OSNN designs. When training the DNN, penalty terms encouraging higher sparsity can be added to the training objective, allowing for the elimination of unneeded $\pmb{\Sigma}$ units while minimizing task performance degradation. Our simulation results indicate that more than 70% of neural connections in our OSNN can be pruned with negligible (<0.2%) accuracy loss when implementing image recognition tasks such as MNIST [41] or FashionMNIST [54]. (Results are provided in Supplementary Note 9). On these datasets, our pruned OSNN can save around 70% of trainable optical components, resulting in ~52% chip area reduction compared to unpruned OSNN.

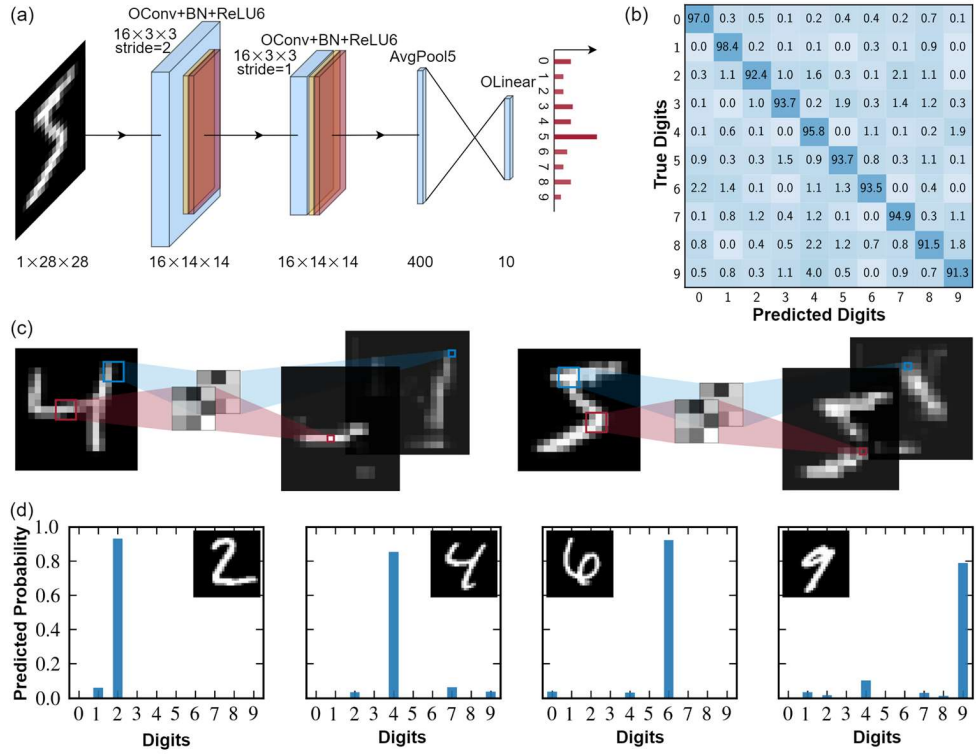

**Fig. 4 Experimental data of digit recognition with the OSNN.** (a) Structure of the CNN, the convolution is realized by OSNN with the im2col approach. The first convolutional layer has one input channel and 16 output channels with a stride of 2. The subsequent convolutional layer has 16 input/output channels with a stride of 1, and the size of the convolutional kernel is 3×3. After adaptive average pooling, we have 5×5×16=400 hidden features, followed by a linear classifier with 10 outputs. (b) The confusion matrix of the trained OSNN on MNIST, showing a measured accuracy of 94.16%. (c) Experimental results of convolving two input images with convolution kernels of size 3×3 in our OSNN. (d) The predicted probability distribution of our OSNN on four selected test digits in the MNIST dataset.

**Computational speed and energy efficiency.** Our OSNN utilizes light to implement MVM operations, which outperforms electronic counterparts in both speed and energy efficiency. Taking into account the delay contributed by high-speed modulators (10 ps) [42,55], photodetectors (10 ps) [56], ADCs (100 ps) [57], and the optical path (43.8 ps), the total delay required to implement a 32×32 MVM can reach ~164 ps, which corresponds to an operating frequency of around 6 GHz . Using the same component library [58], the propagation delay of the optical path in our OSNN is 5.5× less than that of an MZI-based ONN [47], as depicted in Fig. 5(b). The computational speed of OSNN is now constrained by optical-to-electrical (OE) or electrical-to-optical (EO) conversion, but it can be increased further by using all-optical devices as non-linear activation functions [38] (See Supplementary Note 10).

The total power consumption of OSNN for MVM operations is comprised of the power to drive the laser/modulators/photodetectors, the power to set the weight matrix, and the power to drive the ADCs. Numerous energy-efficient active optical components have been developed in recent years. For instance, the silicon microdisk modulator achieves approximately 1 fJ per bit [42]. Maintaining the weight matrix takes less than 2.5 mW per phase shifter in our AMF-manufactured neural chip, which can be decreased to zero by setting weights with phase change materials or nano-opto-electro-mechanical devices [59,60]. Concerning the power consumption of ADCs, despite the availability of high-speed ADCs, the power consumption of ADCs is significantly higher than that of other components. For example, an 8-bit, 40 GSPS ADC consumes 200 mW per channel, while an 8-bit, 10 GSPS ADC consumes 39 mW per channel [57]. In addition, the number of trainable devices in our $k$-point OSNN is only $\mathcal{O}\left(\frac{mn}{k}\right)$, which saves energy for storing and reconfiguring weights. In comparison to ONN architectures designed for general MVM, where the number of programmable devices is around $\mathcal{O}(mn)$ [61] or $\mathcal{O}(\max(m^2, n^2))$ [47], the memory cost of storing and accessing the weight matrix and the energy required to reconfigure corresponding active devices are also reduced by $k$ times. This feature of OSNN will bring considerable energy efficiency improvement when weights need to be reconfigured frequently in large-scale DNNs, where weight loading takes nontrivial hardware cost even with weight-stationary dataflow [62].

**Resolution analysis.** Our OSNN is capable of achieving a high accuracy under low-bit control of optical components. Prior ONN architectures designed for general MVMs require high-precision control of optical devices for parameter mapping to maintain accuracy [47]. Otherwise, we may encounter severe task performance degradation because of large mapping errors [63], which will quickly accumulate as the size of weight matrices or the number of layers increases. Given that the control precision of some energy-efficient photonic tensor cores is only 4 or 5 bits [31], it is necessary to reduce the resolution requirement of ONN architectures and enhance the tolerance of quantization errors. In this study, quantization-aware training is applied to our OSNN to adopt the limited voltage control precision and mitigate the accuracy loss. In experiments, we have shown that ~94% accuracy can be achieved for digit recognition when the precision of the DACs for controlling the phase shifters is around 3-bit (See Supplementary Note 8). What is more, low-resolution device control can also lessen the energy cost for weight storage, access, and reconfiguration [64].

**Robustness.** The robustness of our OSNN is guaranteed by our hardware-aware training framework. Our AI-assisted DPE provides accurate variation modeling of static noises, e.g., process variations, device calibration errors, thermal crosstalk, and non-ideal extinction ratio of modulators. Besides, our noise-injection training algorithm further considers the impacts of dynamic noises, e.g., thermal noises from the laser source and photodetection noises. The robustness of our architecture is evaluated by varying the signal-to-noise-ratio (SNR) of the inputs and the phase drifts of phase shifters in MZI attenuators, and our analysis results are shown in Fig. 5(c). Thanks to our noise injection techniques, our OSNN maintains greater than

90% average inference accuracy even when the standard deviation of input noise and phase drifts reach 0.1 and 0.2, respectively.

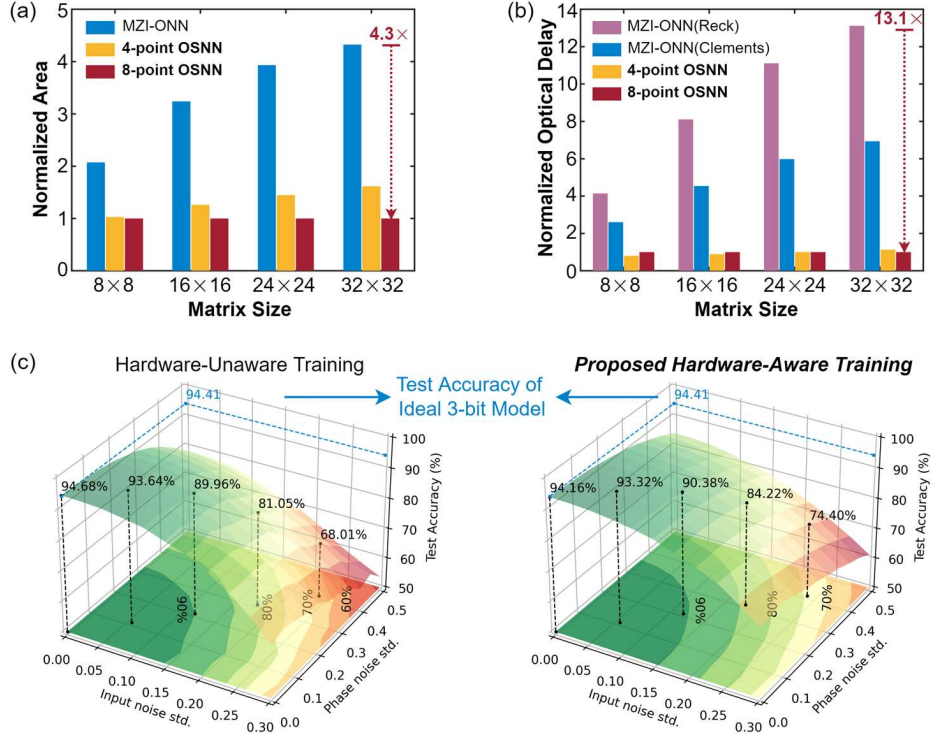

**Fig. 5 Performance analysis of the OSNN.** (a) Normalized area comparison and (b) normalized optical delay comparison between OSNN and MZI-based ONN[11] when implementing weight matrices of different sizes. (c) Robustness comparison between the hardware-unaware training and our proposed hardware-aware training, which includes dynamic noise injection techniques. With noise-awareness, our proposed hardware-aware training flow can effectively boost the noise tolerance of OSNN against various nonideal factors, such as input noses and weight-encoding noises. As a reference, we can achieve 94.41% test accuracy with 3-bit weight programming precision using the ideal transfer-matrix-model of the BPNC but will suffer complete malfunction (~10%) when mapped onto the real chip due to the huge discrepancy between the simulation model and the manufactured chip.

Additionally, the robustness of our OSNN can be enhanced with more reliable optical components and more reliable control circuits [65]. For bandwidth-driven and robustness-driven OSNN design, one can directly select broadband MZI with low temperature sensitivity as a robust variable optical attenuator (VOA). Less robust but more compact or energy-efficient components can also be employed, e.g., ultra-compact MRR modulators with on-chip feedback controls and PCM-based modulators with advanced high-endurance materials.

**Scaling and outlook.** The performance metrics of the OSNN can be further improved in several directions. First, our OSNN is compatible with the majority of the device-level enhancement techniques. For example, by using smaller directional couplers [66], crossings [67], and VOAs [68], the chip area of the OSNN can be optimized, resulting in a competitive computing density of >200 TOPS/mm$^2$ and energy efficiency of ~9.5 TOPS/W (See Supplementary Note 13). Second, massive multiplexing techniques can substantially boost the throughput of our architecture. Because all of the optical components in our architecture can be broadband devices, wavelength-division multiplexing (WDM) techniques can be applied to our architecture: If $k$-wavelength input signals propagate through the chip simultaneously to implement the MVM in parallel, the throughput and the computing density can then be improved by ($k$-1) times over a single-wavelength OSNN. Furthermore, more circuit structures and optical components can be investigated to construct our OSNN. Notably, the BPNC is not the only option to implement $B\Sigma P$. For example, recent work demonstrates that multiport $n$-to-$n$ directional couplers, multimode interference (MMI) couplers, and diffractive cells can be utilized to build unitary matrices [69,70]. They can also be used to build the $B$ and $P$ unit to reduce the chip area. Finally, faster or more energy efficient EO/OE conversion techniques are demanded to improve the computational speed and energy efficiency for data movement between electrons and photons, which currently restricts the performance of optical computing platforms.

**6. Conclusion**

We present a hardware-efficient optical subspace neural network (OSNN) architecture with experimental demonstrations on a silicon photonic programmable butterfly-style photonic-electronic neural chip (BPNC). By exploring optical neurocomputing beyond conventional GEMMs with restricted weight representability, our OSNN consumes up to 7× fewer trainable optical components than prior MZI-based ONN architectures designed for GEMMs. This advantage can be further increased to ~23× using structured circuit pruning strategies with negligible accuracy loss. Our proposed hardware-aware training framework efficiently models the behavior of the OSNN to help reduce control precision requirements, enhance noise robustness, and fully exploit the expressivity in the subspace. The performance of OSNN can be further improved with smaller optical components as well as faster and more efficient EO/OE conversion techniques. Our OSNN pushes the limits of scalability and the robustness of ONNs and creates a new design paradigm for next-generation high-performance AI accelerators with improved hardware efficiency.

*Acknowledgments*

The authors acknowledge support from the Multidisciplinary University Research Initiative (MURI) program through the Air Force Office of Scientific Research (AFOSR), monitored by Dr. Gernot S. Pomrenke.

**Competing interests**

The authors declare no competing interests.

**Data availability**

The data and codes that support the findings of this study are available from the corresponding author upon reasonable request.